\documentclass[a4paper,aps,twocolumn]{revtex4}
\RequirePackage[english]{babel}
\RequirePackage[latin1]{inputenc}
\RequirePackage[T1]{fontenc}
\RequirePackage{mathrsfs}
\RequirePackage{amsmath}
\RequirePackage{amssymb}
\RequirePackage{amsbsy}
\RequirePackage{bm}
\begin{document}
\title{\bf{FROM EXTENDED GRAVITY WITH TORSION-SPIN COUPLING\\
TO RUNNING CONSTANT FOR WEAK-LEPTONIC FORCES}}
\author{Luca Fabbri}
\affiliation{INFN \& Dipartimento di Fisica, Universit\`{a} di Bologna\\
DIME Sez. Metodi e Modelli Matematici, Universit\`{a} di Genova}
\begin{abstract}
In this paper we show how in the most natural extension of gravity with torsion, fermion fields are endowed with running coupling in spinorial interaction, that are shown to reproduce the strength and structure of the leptonic weak forces.
\end{abstract}
\maketitle
\section{Extended Torsion Gravity for Running Weak Coupling}
In any theory of fields the dynamics is expressed by means of differential field equations; and in the theory of relativity all our equations must be covariant under any possible coordinate transformations: what this implies is that covariant derivatives must be defined, a task that is achieved by introducing a specific structure called connection. Additionally there is an ubiquitous principle in physics requiring to deal in the most general situation possible: this means that connections in general are not symmetric in the two lower indices. The antisymmetric part in those two lower indices is thus in general different from zero and it is called Cartan Torsion tensor.

Although there are no \emph{a priori} reasons to remove torsion, there may be \emph{a posteriori} reason to do that linked to the Principle of Equivalence: this reason is based on the idea that if the Principle of Equivalence is to be used to have the unambiguous geometrization of gravity then the vanishing of torsion is to be a sufficient condition \cite{m-t-w}. However in the most comprehensive case in which we want both necessary and sufficient conditions, torsion only needs to be completely antisymmetric \cite{f/1a,f/1b,a-l,m-l}.

Such a geometric underlying background with two fundamental quantities, torsion and curvature, is the best suited to host matter fields, which in general have both spin and energy density. When field equations coupling geometry and matter are defined, then torsion gives the possibility to couple also the spin much in the same way in which curvature provides the way to couple the energy.

The Sciama--Kibble completion of Einstein gravity is given by the field equations
\begin{eqnarray}
&Q^{\rho\mu\nu}=-16\pi kS^{\rho\mu\nu}\\
&G^{\mu}_{\phantom{\mu}\nu}-\frac{1}{2}\delta^{\mu}_{\nu}G
-\lambda\delta^{\mu}_{\nu}=8\pi kT^{\mu}_{\phantom{\mu}\nu}
\end{eqnarray}
for spin and energy densities that in the case of the simplest Dirac matter field are
\begin{eqnarray}
&S^{\rho\mu\nu}
=\frac{i\hbar}{4}\overline{\psi}\{\gamma^{\rho},\sigma^{\mu\nu}\}\psi
\label{spin}\\
&T^{\mu}_{\phantom{\mu}\nu}
=\frac{i\hbar}{2}\left(\overline{\psi}\gamma^{\mu}D_{\nu}\psi
-D_{\nu}\overline{\psi}\gamma^{\mu}\psi\right)
\label{energy}
\end{eqnarray}
with Dirac matter field equations
\begin{eqnarray}
&i\hbar\gamma^{\mu}D_{\mu}\psi-m\psi=0
\label{matterequations}
\end{eqnarray}
as it is well-known. Here $Q^{\rho\mu\nu}$ is the torsion tensor and $G^{\mu}_{\phantom{\mu}\nu\alpha\beta}$ the Riemann tensor of the most general connection containing torsion while its contractions given by the Ricci tensors are given as usual, and finally $D_{\mu}\psi$ is the covariant derivative of the connection with torsion applied to the Dirac field, so that when the torsional contributions are split from within curvatures and derivatives and substituted with the spin density of the spinor field, the entire set of field equations reduces to
\begin{eqnarray}
&i\hbar\gamma^{\mu}\nabla_{\mu}\psi
-3\pi k\hbar^{2}\overline{\psi}\gamma_{\mu}\psi\gamma^{\mu}\psi-m\psi=0
\end{eqnarray}
that is the one we would have had in the torsionless case but supplemented by fermionic potentials of the Nambu--Jona-Lasinio form, with $\nabla_{\mu}\psi$ being the covariant derivative of the connection without torsion for the Dirac field.

Now these field equations might have also been obtained through the variation of the lagrangian density given in terms of the Dirac least-order derivative lagrangian density as it is widely known, that is a Lagrangian of the type $G$; but one can get generalizations for which not $G$ but a generic function $f(G)$ is considered.

In this case, what happens is that the field equations for torsion and gravity are much more complicated, but nevertheless the field equations for matter remains very similar indeed. With the introduction of the function $\phi=f'(G)$ the decomposed Dirac matter field equations are shown to have the following expression
\begin{eqnarray}
&i\hbar\gamma^{\mu}\nabla_{\mu}\psi
-\frac{\hbar^{2}}{\phi}\overline{\psi}\gamma_{\mu}\psi\gamma^{\mu}\psi-m\psi=0
\end{eqnarray}
which are the Nambu--Jona-Lasinio equations but now with a running coupling constant \cite{Fabbri:2010pk}.

In the case we would have studied not the single but the couple of fermions, such as the case of two leptons, the same field equations would have read
\begin{eqnarray}
&i\gamma^{\mu}\nabla_{\mu}e
-\frac{\hbar^{2}}{\phi}\overline{e}\gamma_{\mu}e\gamma^{\mu}e
-\frac{\hbar^{2}}{\phi}\overline{\nu}\gamma_{\mu}\nu\gamma^{\mu}\gamma e-me=0
\label{1}\\
&i\gamma^{\mu}\nabla_{\mu}\nu
-\frac{\hbar^{2}}{\phi}\overline{e}\gamma_{\mu}\gamma e\gamma^{\mu}\nu=0
\label{2}
\end{eqnarray}
which can be Fierz rearranged into the equivalent form
\begin{eqnarray}
\nonumber
&i\gamma^{\mu}\nabla_{\mu}e+\frac{\hbar^{2}}{\phi}2(\cos{\theta})^{2}\overline{e}\gamma e\gamma e
+q\tan{\theta}Z_{\mu}\gamma^{\mu}e-\\
&-\frac{g}{2\cos{\theta}}Z_{\mu}\gamma^{\mu}e_{L}
+\frac{g}{\sqrt{2}}W^{*}_{\mu}\gamma^{\mu}\nu-YHe-me=0
\label{electron}\\
&i\gamma^{\mu}\nabla_{\mu}\nu+\frac{g}{2\cos{\theta}}Z_{\mu}\gamma^{\mu}\nu
+\frac{g}{\sqrt{2}}W_{\mu}\gamma^{\mu}e_{L}=0
\label{neutrino}
\end{eqnarray}
upon definition of the bosons, being them scalars
\begin{eqnarray}
&H=\frac{\hbar^{2}}{\phi Y}2(\cos{\theta})^{2}\overline{e}e
\label{Higgs}
\end{eqnarray}
or vectors
\begin{eqnarray}
&\!Z_{\mu}\!=\!\frac{\hbar^{2}}{\phi}\!\frac{2\cos{\theta}}{g\left(2\sin{\theta}\right)^{2}}
\!\!\left[2\left(\overline{e}_{L}\gamma_{\mu}e_{L}\!-\!\overline{\nu}\gamma_{\mu}\nu\right)
\!-\!(2\sin{\theta})^{2}\overline{e}\gamma_{\mu}e\right]
\label{neutral}\\
&W_{\mu}\!=\!\frac{\hbar^{2}}{\phi}\frac{\sqrt{2}\left[1-(2\sin{\theta})^{2}\right]}{g\left(2\sin{\theta}\right)^{2}}
\left(2\overline{e}_{L}\gamma_{\mu}\nu\right)
\label{charged}
\end{eqnarray}
as a direct calculation would show. If the energy is low enough not to probe the internal structure of these bosons, so that they may be taken as approximately structureless, then the leptonic field equations are those we would have had in the torsionless case but with weak forces among leptons as in the standard model and with a coupling running as the energy of the weak forces \cite{Fabbri:2010ux,Capozziello:2012gw}.

This may well be a coincidence, but the fact that such general torsion fields can reproduce the peculiar structure of weak forces and that such natural $f(G)$ extensions can give rise to the typical Fermi energy scale is certainly intriguing enough to pursue further investigations: for instance, would it be possible to calculate for these weak composite mediators important quantities such as effective masses? And other quantities such as branching ratios? Given that here the Higgs field is composite in terms of leptonic fields bound together by torsion, what are the discrepancies with the Higgs sector of the standard model? Some of these questions are already work-in-progress, for some others we are still at the beginning, and in general it is too early to consider this approach as a complement to the standard model.

But one thing that is certain is that we cannot underestimate the potential torsion in extended gravity may have for particle physics.

\end{document}